\documentclass{aa}  

\usepackage{graphicx}
\usepackage{color}
\usepackage{txfonts}
\usepackage[]{natbib}
\newcommand{\healpix}{\textsc{healpix}}
\newcommand{\Nside}{\textit{Nside}}
\newcommand{\class}{\textsc{class}}

\begin{document}

   \title{Angular two-point correlation of NVSS galaxies revisited}

   \author{Song Chen\thanks{\email{songchen at physik.uni-bielefeld.de}}
          \and
          Dominik J. Schwarz\thanks{\email{dschwarz at physik.uni-bielefeld.de}} 
          }

   \institute{Fakult\"{a}t f\"{u}r Physik, Universit\"{a}t Bielefeld, Postfach
100131, 33501 Bielefeld, Germany              
          }

   \date{Preprint online version: July 8th, 2015 }

\abstract{We measure the angular two-point correlation and angular power spectrum from the 
NRAO VLA Sky Survey (NVSS) of radio galaxies.
They are found to be consistent with the best-fit cosmological model from the Planck analysis, 
and with the redshift distribution obtained from the Combined EIS-NVSS Survey Of Radio Sources (CENSORS).  
Our analysis is based on an optimal estimation of the two-point correlation function and makes 
use of a new mask, that takes into account direction dependent effects of the observations, sidelobe
effects of bright sources and galactic foreground. We also set a
flux threshold and take the cosmic radio dipole into account. The latter turns out to be 
an essential step in the analysis. This improved cosmological analysis of the NVSS
emphasizes the importance of a flux calibration that is robust and stable on large angular scales for future 
radio continuum surveys.}


\maketitle


\section{Introduction}

Continuum radio galaxy surveys can probe large cosmic volumes. The mean redshift 
of observed radio galaxies is significantly higher than the mean redshift of modern optical surveys. 
Optical surveys allow us to obtain photometric or even spectroscopic redshift measurements, which 
is not the case for continuum radio galaxy surveys. Nevertheless, the information contained 
in continuum radio galaxy surveys can be useful for probing cosmological models. An
example is the investigation of non-Gaussianity via angular two-point correlations. 
In this work we revisit some aspects of the cosmological 
analysis of continuum radio galaxy surveys.

Most of our current understanding of cosmology relies on 
observations of the cosmic microwave background (CMB). 
Planck satellite data extended our understanding to the 
high multipole CMB, and allowed us to successfully measure the cross-correlation between the 
large-scale structure and the CMB through the integrated Sachs-Wolfe (ISW) and lensing effects.
In the study of the ISW effect, large-scale radio continuum catalogues play an important role. 
For the Planck analysis the NRAO VLA Sky Survey (NVSS) \citep{Condon1998} has been 
utilized \citep{planckISW}. 

Radio galaxy surveys, such as the NVSS, have large sky coverage and extend to redshifts well beyond unity. 
Their angular two-point correlation function and angular power spectrum are useful probes of large-scale 
cosmology. With new radio continuum surveys, containing between 10 million to a few billion objects, 
such as from the Low Frequency Array  (LOFAR)\footnote{URL: www.lofar.org}, 
the Australian Square Kilometre Array Pathfinder (ASKAP)\footnote{URL: www.atnf.csiro.au/projects/askap/}, the 
South African MeerKAT radio telescope (MeerKAT)\footnote{URL: http://www.ska.ac.za/meerkat/}
or the Square Kilometre Array (SKA)\footnote{URL: www.skatelescope.org}, more complete and precise 
radio point source catalogues will be available. These surveys will allow us to probe several interesting 
aspects of cosmology (for a recent review see \cite{Jarvis:2015asa}), including accurate 
investigations of the largest cosmic structures. 

For cosmological analysis the fundamental observable from a survey is the surface density 
of sources above a given flux density limit, i.e.~the integral of the differential number count per solid angle 
per flux density interval as a function of position on the sky. A study that includes all effects of the 
surface density at linear order in cosmological perturbation theory has been presented recently 
\citep{Maartens:2012rh,Chen:2014bba}. These effects could be modified by local effects of large structures, 
e.g.~by local voids \citep{Rubart:2014lia}. Radio continuum surveys are also a good candidate to further 
test several CMB anomalies 
\citep{WMAP7-anomalies,CHSS-review,Naselsky:2011jp,Planck-R1-XXIII,Copi:2015a,Copi2015b,planck:2015}.  

\begin{figure*}
\includegraphics[width=\linewidth]{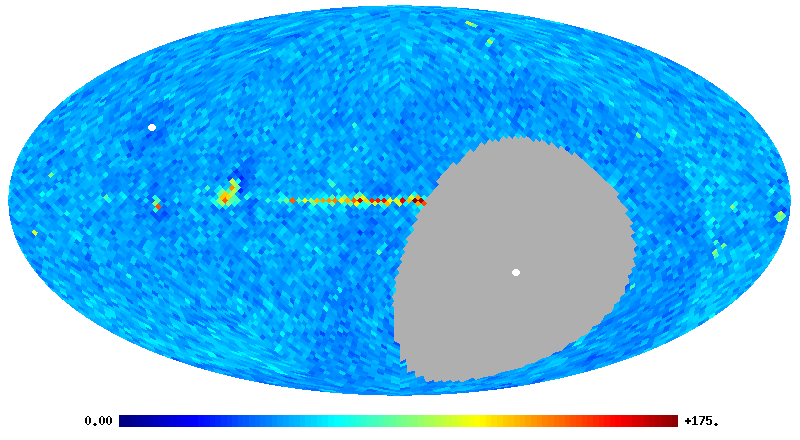}
\caption{ 
Surface density of the NVSS source catalogue in galactic coordinates in Mollweide projection, shown at 
\healpix{} resolution $\Nside =32$. The colour bar shows the surface density in units of number of objects per 
square degree. Here we include all objects contained in the catalogue. The white 
dots indicate the the positions of the celestial poles.}
\label{fig:full}
\end{figure*}

Previously, \cite{Blake2002} explored the angular two-point correlation of the NVSS catalogue 
at small angular separation ($\theta < 10^\circ$). In addition, they also measured the angular power spectrum of 
the same catalogue \citep{Blake2004a}, and found that their result seems to be incompatible with a high 
redshift population of radio galaxies. \cite{Xia:2010yu} suggested that the clustering at 
$1^\circ \leq\theta \leq 8^\circ$ is dominated by some non-Gaussian component. They quote the  
non-Gaussianity parameter $25 < f_{\rm nl} < 117$ at $2\sigma$. 
The angular power spectrum was also tested more recently in \cite{Marcos-Caballero:2013yda}, 
which also find an excess of power at large scales, but $-43 < f_{\rm nl} < 142$ at $2\sigma$.
\cite{refId0} investigated in detail the two-point cross-correlation between the 
NVSS catalogue and the CMB temperature anisotropies from the Wilkinson Microwave Anisotropy Probe (WMAP) 
in angular and multipole space. He did not find any evidence for the cross-correlation
in the range $l \in [2,10]$ where, according to the simulation 50\% of the signal is expected. He concluded that 
the large-scale clustering excess, which he found in the NVSS catalogue, is unlikely to be caused by 
contaminants or systematics since it appears to be independent of the flux density threshold. 

In this work, we explore the full range of the angular two-point correlation function ($0^\circ < \theta < 180^\circ$) 
with a focus on large angular scales ($\geq 1^\circ$) and compare it with the flat $\Lambda$ cold dark matter 
($\Lambda$CDM) model. The focus of our investigation lies on how to prepare a catalogue from a 
real all-sky radio continuum survey for cosmological analysis, rather than the estimation of cosmological 
parameters. All previous analyses are based on masking of the galaxy by a constant latitude cut.
Moreover  the cosmic radio dipole \citep{1984MNRAS.206..377E,Blake:2002gx} has not 
been accounted for in previous works. To extract the auto-correlation signal from the catalogue in a reliable way, 
we propose a more sophisticated NVSS mask and subtract the cosmic radio dipole. Based on 
sidelobe contamination estimates and noise properties of the NVSS catalogue, we provide a mask that 
leaves us with a sky coverage of 64.7\% at a minimal specific flux density of $15$~mJy.

Our work is structured as follows. In section \ref{sec:1}  we review the properties of the NVSS catalogue and 
explain our masking method. In section \ref{sec:2}, we review the theoretical angular two-point 
correlation function based on the $\Lambda$CDM model. We present the angular 
two-point correlation obtained by means of the Landy-Szalay estimator using our mask in section 
\ref{sec:3}. In section \ref{sec:4} we discuss our results, followed by the conclusion in section \ref{sec:6}. 


\section{\label{sec:1} NVSS catalogue}

The NRAO VLA Sky Survey \cite{Condon1998} used the D and DnC antenna configurations of the 
Very Large Array (VLA). The survey was carried out between 1993 and 1997. Continuum intensity and 
linear polarization 
images at $1.4$ GHz were obtained, covering the whole northern and part of the southern sky at 
declinations $\delta > - 40^\circ$. The D configuration of the VLA was used to cover the sky from 
$\delta = -10^\circ$ to $\delta = 78^\circ$, the rest was filled in by means of the DnC configuration. 
The obtained images have on average a FWHM resolution $\theta_{\rm FWHM}=45''$, which is significantly 
larger than the median angular size of faint extragalactic sources (around $10''$). 
This survey design sacrifices resolution to achieve high surface brightness, which is needed to achieve 
flux-limit completeness. 

Since the NVSS point-source response is much larger than the median angular size of extragalactic sources,
most of the information on the NVSS total intensity images is represented well by elliptical Gaussian fits.
The fitted parameters formed the NVSS catalogue of sources (not all of which are individual sources). 
The NVSS catalogue contains almost 2 million discrete sources with flux density above $2.5$ mJy. 
Below we use the source positions in right ascension $\alpha$ and declination $\delta$, the 
integrated flux density $S$, and their errors.

The NVSS catalogue contains several artificial effects \citep{Ho:2008bz}: First, the catalogue shows a 
configuration effect, which is easily seen in figure \ref{fig:full} as steps in surface density at 
$\delta \sim 78^\circ$ and $\delta \sim - 10^\circ$, which are the borders between the D and DnC 
configurations on the map.
Second, the sidelobes of the bright sources can obscure the presence of weak nearby sources. 
They are not completely removed by cleaning. The NVSS local dynamic range of the total intensity images is 
about $1000$ to $1$. Thus, sources closer than about $0.6^\circ$ to a bright source of flux density $S$ and 
fainter than $10^{-3}S$ can only be caused by sidelobes \citep{Condon1998}.

In addition to these two unphysical effects, there is also a foreground of point sources mainly from the 
Milky Way (see figure \ref{fig:full}) and smooth components of radiation from the galaxy itself. 
These contaminations can influence the VLA noise temperature, and further change the 
completeness in certain areas on the sky.

All these effects need to be treated carefully, otherwise the statistical analysis based on the 
contaminated map will be biased. Here we choose to apply a lower flux density threshold and 
to mask the catalogue to address these issues.  

\subsection{Lower flux density threshold}

Two different VLA configurations (D and DnC) have been used to compile the NVSS catalogue. 
The VLA C configuration is less compact and has a resolution of $15''$, which is better than $45''$ 
of the D configuration. 
The DnC configuration is a hybrid configuration in which the antennas on the east and west arms are in D 
configuration, but those on the north arm are in C configuration to enhance their view of sources at low 
elevation. Using the DnC configuration changes the synthesized beam and the resolution of declination
with respect to the D configuration. This shifts the brightness sensitivity and completeness between the 
parts of sky observed with different configurations. 

The source catalogue is derived from intensity images, therefore it is brightness sensitivity limited. 
Apparently, the D configuration has higher brightness 
sensitivity than the DnC configuration. Thus a more complete catalogue for the part 
of the sky observed by means of the D configuration is expected.

The completeness shift of the DnC configuration sky with respect to the D configuration sky can be considered as 
a faint source selection based on the position angle, noise, confusion and cataloguing procedures.
It is not clear whether this selection also introduces a tension in source distribution between the two parts 
of the sky. Therefore, it is safe to either use one configuration alone (which would reduce the sky coverage by a 
significant amount), or to choose a higher flux density threshold for the cosmological analysis (which 
reduces the source density and increases the shot noise). 

\begin{figure}
  \includegraphics[width=\linewidth]{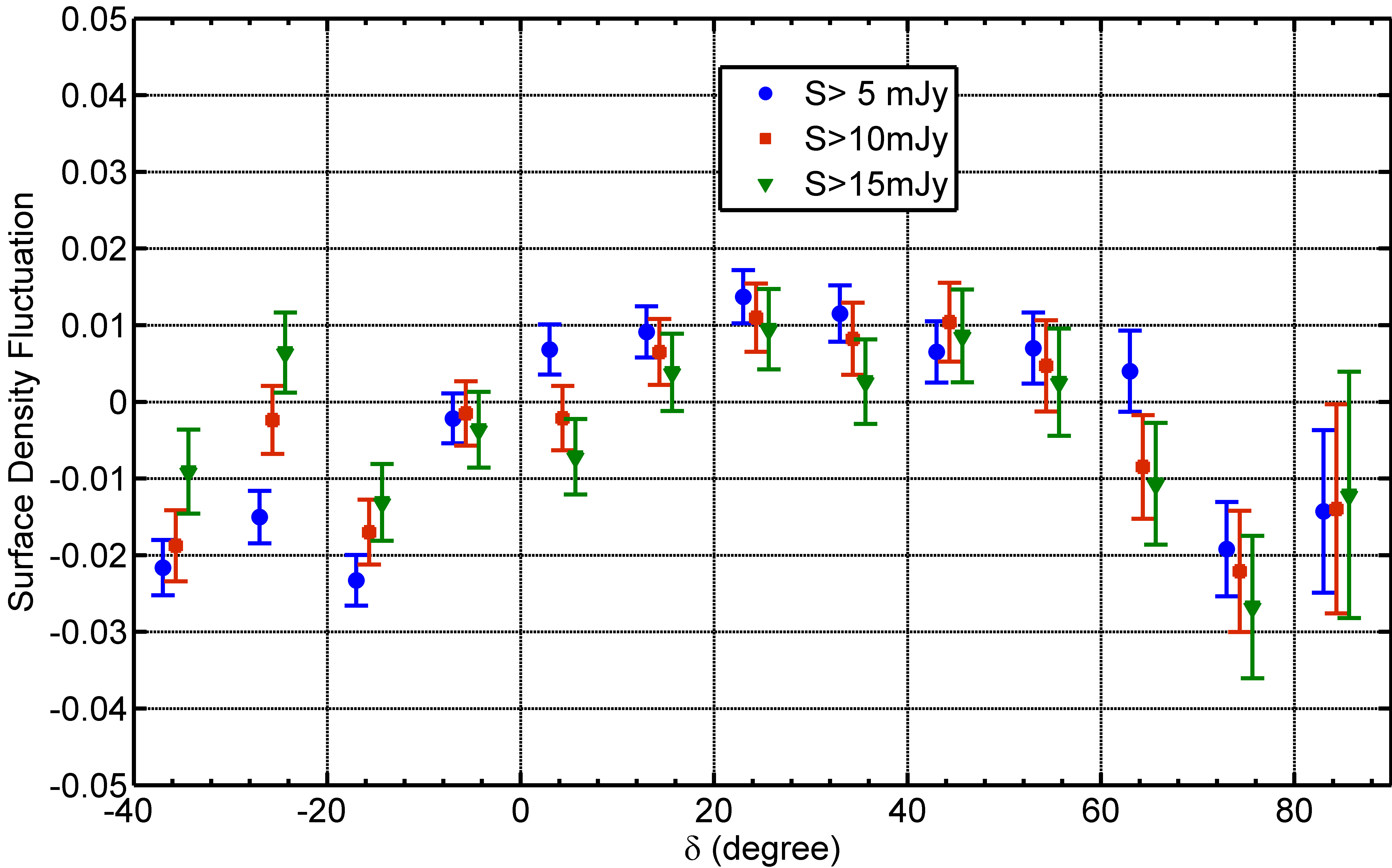}
  \caption{Surface density fluctuation, $\Delta\sigma/\overline{\sigma}$, of the NVSS catalogue as a function of 
  declination for several flux density thresholds. The error bars assume Poisson distributed source 
  counts and we mask the region $|b| \leq 5^\circ$. For clarity, the horizontal positions of the 
  $S > 5$ mJy and $S> 15$ mJy data points are slightly offset.}
  \label{fig:strip-flux}
\end{figure}

\cite{Blake2002} argue that this configuration effect is only significant at flux densities 
$S<10$ mJy. In figure \ref{fig:strip-flux} we plot the surface density fluctuation, $\Delta\sigma/\overline{\sigma}  
= \sigma/\overline{\sigma} - 1$, as a function of declination. The dependence of the surface density fluctuation 
on declination resembles the theoretical rms noise level of the NVSS catalogue \citep{Condon1998}, which 
is discussed further in section \ref{sec:4}. It can be seen that the declination dependence of the 
surface density is less than $2.5\%$ and the mean value of the DnC configuration is clearly lower 
than that of the D configuration. This configuration effect is more pronounced at lower thresholds.
 
This finding is also supported by a $\chi^2$-analysis testing for deviations from isotropy in declination
\begin{equation}
\chi^2 \equiv \sum_{i = 1}^{N_{\rm bin}} \frac{\left({\Delta\sigma_i\over \bar\sigma}\right)^2}
{\left(\delta\left[{\Delta\sigma\over \bar\sigma}\right]_i\right)^2} = 
\sum_{i = 1}^{N_{\rm bin}} \Omega_i \frac{\left(\Delta\sigma_i\right)^2}{\sigma_i}, 
\end{equation}
where $\Omega_i$ denotes the solid angle of the $i$th declination strip. The results of this test are shown in 
table~\ref{tab:NVSS-strip-chi2}. 

\begin {table}
\begin{center}
\begin{tabular}{c | c | c | c | c  }
$S >$  & $5$ mJy & $10$ mJy & $15$ mJy & $25$ mJy \\  \hline
$\chi^2$  & 160.1 & 59.7 & 30.7 & 21.5
\end{tabular}
\end{center}
\caption {$\chi^2$-values testing for the isotropy in declination of the NVSS surface density after 
masking the region $|b| \leq 5^\circ$ and for 13 degrees of freedom.}
\label{tab:NVSS-strip-chi2}
\end{table}

In this work, we choose flux density thresholds of $15$ mJy and $25$ mJy for cosmological analysis 
for the following reasons: The study of the configuration effect suggests that there is a significant 
dependence on declination. As is shown above, a flux density threshold of $15$ mJy or $25$ mJy 
reduces the value of $\chi^2$ dramatically compared to the situation with lower thresholds. As we do not 
expect a perfect agreement with isotropy, it is not justified to raise the threshold even higher.  

Our choice is supported by studies of the cosmic radio dipole from NVSS, which is one to two 
orders of magnitudes larger than the higher multipole moments. The previous NVSS dipole 
measurements show that for thresholds below $15$ mJy, the dipole direction differs significantly 
from the CMB dipole direction, and the reduced $\chi^2$ of the quadratic dipole estimator increases 
significantly \citep{Blake:2002gx,Gibelyou:2012ri,Rubart:2013tx,Rubart:thesis}. 
 
No cut-off at high flux densities is introduced as there is only a small number of 
sources at high flux densities and thus they only play a subdominant role. We also expect them to be less 
affected by calibration systematics. 

\subsection{Masking strategy}

Nevertheless, we have to mask regions around the brightest sources in order to minimize contamination from 
their sidelobes. In the analysis of \cite{Ho:2008bz} all sources above $2.5$ Jy have been masked by 
a disk with a radius of $0.6^\circ$. However, not all sidelobes appear as spurious entries in the catalogue, a 
uniform masking strategy will erase all the information from bright sources, since all sources above the cut 
threshold are masked. In \cite{Blake2002} a list of 22 masks around bright galaxies was compiled based on 
a visual inspection of the survey. 

\begin{figure}
  \includegraphics[width=\linewidth]{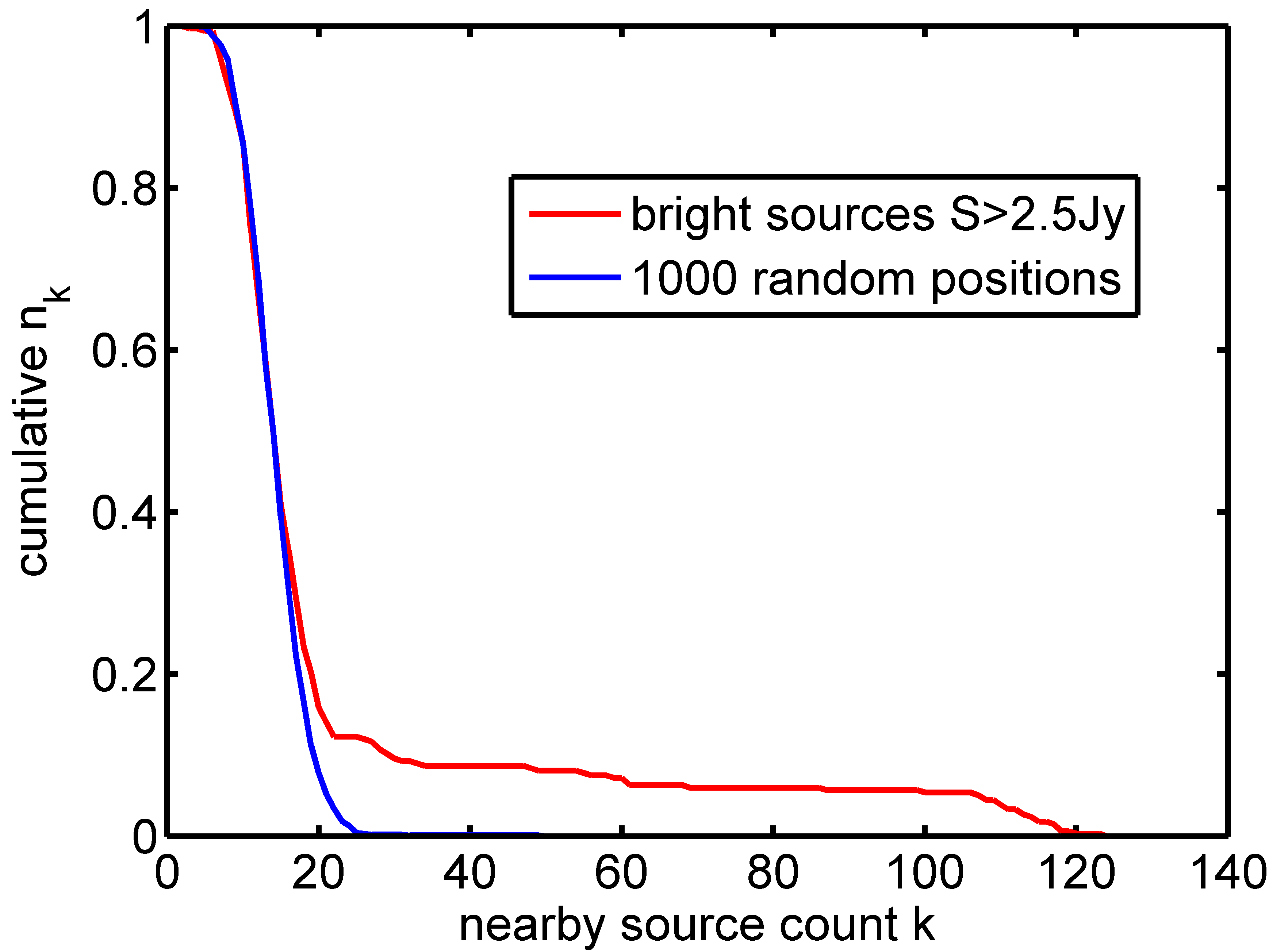}
  \caption{ Cumulative distribution of nearby source counts
  (disks with radius $0.6^\circ$). The red line corresponds to nearby 
  source counts around bright radio galaxies with $S>2.5$ Jy. The blue line corresponds 
  to 1000 randomly picked positions outside the galactic plane ($|b|>5^\circ$). The maximum of 
  randomly picked counts results in 49 nearby sources.}
  \label{fig:hist15}
\end{figure}

Here, we introduce an automatic bright source selection method.
We count the number of nearby sources (within $0.6^\circ$) 
with $S>15$ mJy of the brightest sources whose 
flux densities are $S>2.5$ Jy. The corresponding histogram is shown in Fig.~\ref{fig:hist15}.

\begin{figure}
  \includegraphics[width=\linewidth]{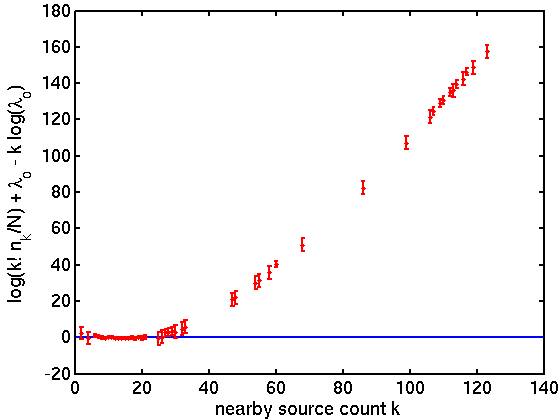}
  \caption{Poissonness plot of the nearby source counts. $\lambda_o$ is the mean nearby count over 
  the survey area. $N$ is the total number of sources with $S>2.5$ Jy. The solid horizontal line corresponds to 
  a perfectly Possion distributed nearby source count. The error bars denote the $99\%$ confidence levels.}
  \label{fig:poisson-15}
\end{figure}

We assume that the number of sources in a fixed solid angle follows a Poisson distribution, and use the 
so-called Poissonness plot \citep{possionplot} to identify the histogram bins that significantly deviate from 
a Poisson distribution, where we assume that such a deviation is caused by the sidelobe contamination of 
bright sources. The idea of the plot is to consider a simple variable substitution, which transforms the exponential 
Poisson distribution function to a linear function.
We find that $49$ sources fail the Poissonness test at the $99\%$ confidence level 
(Fig.~\ref{fig:poisson-15}). It is worth clarifying that `clean' regions containing bright sources that, by chance, 
contain the same amount of sources as 'dirty' regions are also excluded. We also verify that for most 
of the bright sources the cumulative nearby source distribution is in good agreement with the 
cumulative distribution of randomly picked nearby source counts, which justifies the inclusion of many of 
the bright sources in our analysis.  

\begin{figure}
  \includegraphics[width=\linewidth]{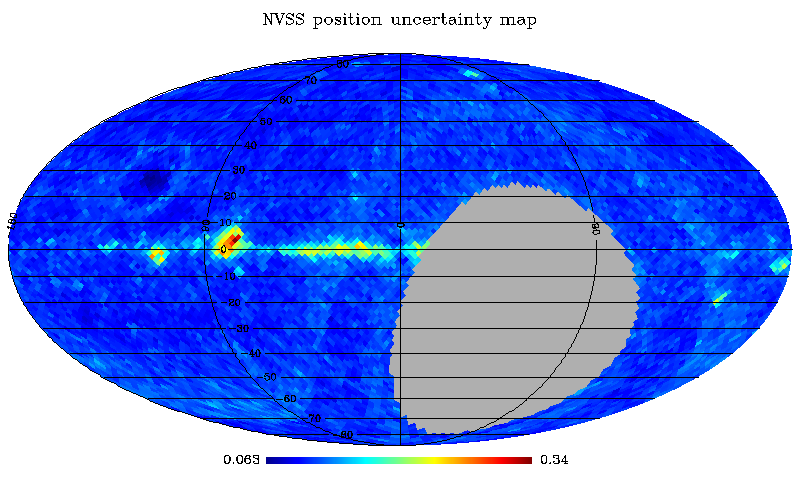}
  \caption{ NVSS position uncertainty map, relative to the mean beam width $\theta_{\rm FWHM} = 45''$.}
  \label{fig:positionRMS}
\end{figure}

We now turn to the issue of noise and confusion. According to \cite{Condon1998}, the rms position 
uncertainty $\sigma_{\rm pos}$ is
\begin{equation}
 \sigma_{\rm pos} \propto \frac{\sigma_{\rm b} \theta_{\rm FWHM}}{2 S_{\rm p}},
\end{equation}
where $S_{\rm p}$ is the peak flux density and $\sigma_{\rm b}$ is the rms brightness fluctuation (noise and confusion).
Our idea is to use $\sigma_p$ to trace $\sigma_{\rm b}$. \cite{Condon1998} point out that for flux 
densities below $15$~mJy, the rms position uncertainty is dominated by noise. Accordingly, we create a 
position uncertainty map (Fig.~\ref{fig:positionRMS}) by averaging the position uncertainties
\begin{equation}
 \sigma_{\theta}\equiv \sqrt{\frac{\sigma_{\delta} \sigma_{\alpha}}{ \theta^2_{\rm FWHM}} \sin(\frac{\pi}{2}-\delta)}
\end{equation}
for all point sources in a pixel whose flux density is smaller than $15$~mJy. 
We note that the dominant sources are those with low flux density. The map is constructed using 
the \healpix{} package with the pixel size fixed by $\Nside = 32$.

The resulting position uncertainty map is shown in Fig.~\ref{fig:positionRMS}. 
Which clearly shows that the position uncertainty follows the theoretical rms noise level which, outside of the 
galactic plane, increases away from the zenith owing to pickup of ground radiation, atmospheric emission, 
ionospheric effects and uv-projection. The galactic plane and nearby nebulas are also easy to identify. 
We also employ a 5\% pixel cutoff for the pixels with highest $\sigma_{\theta}>0.132$. 
In addition, we mask galactic radio sources by excluding all sources with galactic latitude $|b|\leq 5^\circ$.

\begin{figure*}
  \includegraphics[width=\linewidth]{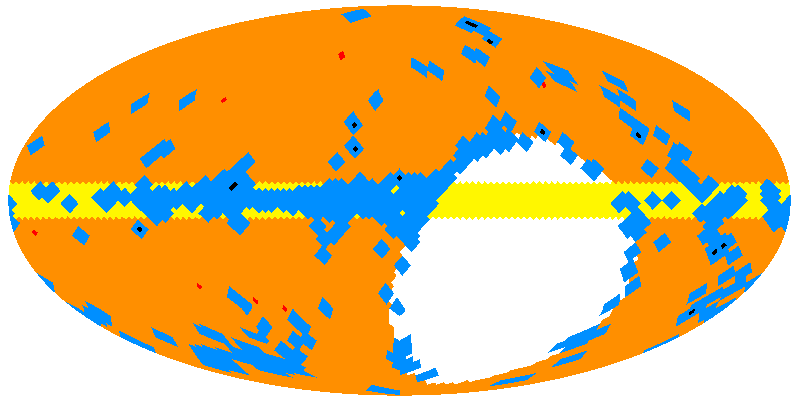}
  \caption{NVSS65 mask.  The orange region makes up $64.7\%$ of the sky. Yellow pixels are close to the 
  galactic plane $|b|\le 5^\circ$ and are excluded to suppress galactic point sources and foregrounds. 
  Blue pixels are excluded owing to large position errors. Red and black pixels contain bright sources with 
  significant sidelobe effects;  black pixels also overlap with pixels with high position error.}
  \label{fig:NVSS-mask}
\end{figure*}

To sum up, at $15$ mJy and $25$ mJy thresholds we mask all pixels in the neighbourhood of 49 selected 
bright sources. Additional pixels and their neighborhood with highest mean position uncertainty 
$\sigma_{\Omega}$ and galactic sources with $|b|\leq 5^\circ$ are masked as well. 
In the following we call this the NVSS65 mask\footnote{ The NVSS65 mask is public available as a source file on the arXiv $1507.02160$.}. It is shown in Fig.~\ref{fig:NVSS-mask}. In total,
approximately $64.7\%$ of the sky is left. The total number of objects after applying the NVSS65 mask at 
$15$ and $25$ mJy is shown in Table~\ref{tab:NVSS-chi2}. 

\begin{figure*}
  \includegraphics[width=\linewidth]{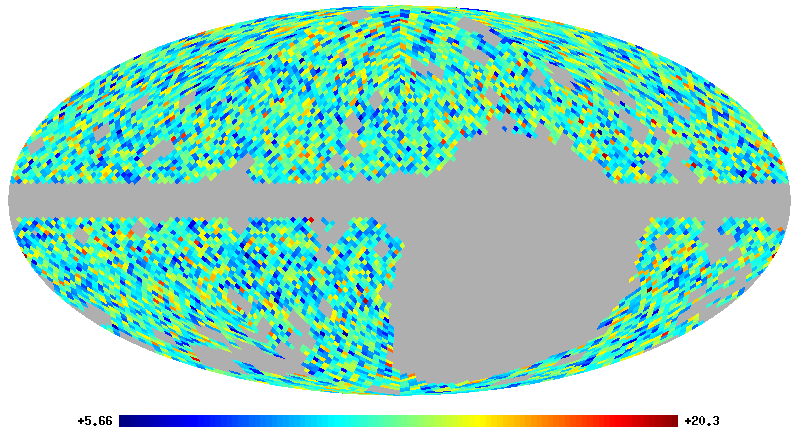}
  \caption{ Surface density of the NVSS source catalogue for a flux density threshold of $S>15$ mJy and applying 
  the NVSS65 mask, shown in galactic coordinates at pixel size \Nside=32. The colour bar shows the 
  surface density $\sigma$ in units of number of objects per square degree.}
  \label{fig:NVSS-mask-15mJy}
\end{figure*}

\section{\label{sec:2} Angular two-point correlation: theoretical expectation}

The angular two-point correlation function $w(\theta)$ is a powerful tool for measuring the projected 
large-scale structure distribution of the Universe. It is defined as the joint probability $\delta P$ 
of finding galaxies in both of the elements of solid angle $\delta\Omega_{1}$ and $\delta\Omega_{2}$ 
separated by an angle $\theta$  \citep{Peebles1980},
\begin{equation}
 \delta P= \bar\sigma^2\delta\Omega_{1}\delta\Omega_{2}[1+w(\theta)].
\end{equation}
At linear order, the full relativistic expression of the surface density of sources above flux density $S_t$ is 
\citep{Chen:2014bba}, 
\begin{eqnarray}\label{eq:result1}
\sigma(>S_t) &=&  \int_{S_t}^\infty \frac{{\rm d}S_{\rm o}}{S_{\rm o}}   \frac{a^3 r_{\rm o}^3 \bar{n}_{\rm phy}}{2+(\bar{\alpha}+1)r_{\rm o}\mathcal{H}} [1+\delta_n+3\psi+\Delta E \nonumber \\
&&+V'^ie_i^r +2\frac{\delta r}{r_{\rm o}}+\frac{\partial \delta
r}{\partial r_{\rm o}}-2\kappa_{\rm g}],
\end{eqnarray}
where $S_{\rm o}$ is the observed flux density, $a$ is the scale factor, $r_{\rm o}$ is the inferred comoving distance, 
$\bar{\alpha}$ is the mean source spectrum index, $\mathcal{H}$ is the conformal Hubble parameter, and 
$\bar{n}_{\rm phy}$ is the physical mean number density. Inside the brackets, the most important contribution 
comes from the number density fluctuation $\delta_n$. The other terms describe  
volume distortion via the scalar metric perturbations $\psi$ and $E$; light cone projection 
involving the projected source velocity $V'^ie_i^r$; radial distance fluctuation 
$2{\delta r}/{r_{\rm o}}+{\partial \delta r}/{\partial r_{\rm o}}$, which lead to a fluctuations of flux density; and 
lensing convergence $\kappa_{\rm g}$. We only take the dominant leading term, 
i.e.~physical number density fluctuation, into account and neglect all other effects. This is justified 
since the statistical and systematic fluctuations from the NVSS catalogue are well above the expected size of 
the subleading linear effects.

The luminosity and density evolution of radio galaxies is significant. 
Therefore, we turn to redshift space, 
\begin{eqnarray}
 \sigma(>S_t)&\approx&c  \int {\rm d}z  \frac{a^3 r_{\rm o}^2}{ H} \bar n_{\rm phy}[1+\delta_n]\\
 &=&\int {\rm d}z \frac{{\rm d} \bar{\sigma}}{{\rm d}z}[1+b \delta] \nonumber, 
\end{eqnarray}
where $b = b(z)$ denotes the bias of radio galaxies and $\delta$ is the matter density contrast in synchronous 
comoving coordinates.   

We use the Combined EIS-NVSS Survey Of Radio Sources (CENSORS) \citep{Brookes:2008fw} to model the 
redshift distribution of the NVSS catalogue. CENSORS contains all the NVSS sources above $7.2$ mJy that are 
within a patch of 6 deg$^2$ in the ESO Imaging Survey (EIS). Following \cite{Marcos-Caballero:2013yda}, 
we choose the gamma function redshift distribution, 
\begin{equation}
 \frac{{\rm d} \bar{\sigma}}{{\rm d}z}={\cal N} \left( \frac{z}{z_{\rm 0}}\right)^\beta \exp\left(-\beta \frac{z}{z_{\rm 0}}\right) \;,
\end{equation}
with ${\cal N}$ denoting a normalization factor that is irrelevant for the final result and the best-fit parameters 
to CENSORS data given by $z_{\rm 0}=0.53^{+0.11}_{-0.13}$ and $\beta=0.81^{+0.34}_{-0.32}$. However, 
these numbers should be treated with care as the redshift distribution is only based on 149 galaxies. 

Owing to the broad shape of the luminosity function, the radio galaxy redshift distribution shows only a 
weak dependence on the flux threshold \citep{deZotti2010,Blake2004a}. This argument holds true, 
because for $S > 1$ mJy  
the differential number counts are dominated by active galactic nuclei, i.e.~the flux 
density threshold is well above the flux density of a typical star forming galaxy.

For a statistically isotropic distribution of radio galaxies the angular two-point correlation $w(\theta)$ can 
be determined from the angular power spectrum $C_l$, 
\begin{equation}\label{eq:cl2w}
w(\theta)=\frac{1}{4\pi}\sum_l (2l+1) C_l P_l(\cos\theta)\;,
\end{equation}
where $P_l(x)$ are Legendre polynomials.
The angular power spectrum $C_l$ of the surface density fluctuation can be calculated from the 
underlying matter power spectrum,
\begin{equation}
 C_l= 4\pi \int_0^\infty  \frac{{\rm d}k}{k} P(k) T(k)^2 f(l,k)^2, 
\end{equation}
with $P(k)$ denoting the primordial power spectrum of density fluctuations, $T(k)$ the matter transfer function and 
\begin{equation}
 f(l,k)=\int \frac{{\rm d}\bar{\sigma}}{{\rm d}z} b(z) j_l[ck\eta(z)] D(z) {\rm d}z, 
\end{equation}
where $({\rm d} \bar{\sigma}/{\rm d}z)$ is the differential number density of sources at flux density above 
$S_t$, $b(z)$ denotes the bias, $j_{\ell}(x)$ is a spherical Bessel function of first kind, $\eta = \eta(z)$ 
denotes conformal time, and $D(z)$ is the growth factor. 

For the galaxy bias, we use the second-order polynomial from the Planck 2015 analysis of the integrated Sachs-Wolfe effect \citep{planckISW15}, which is an approximation of the Gaussian bias evolution model of 
\cite{Xia:2010yu}:  
\begin{equation}
\label{bias}
b(z)=0.9\left[1+0.54(1+z)^2\right].
\end{equation}
Recently, a detailed analysis of the biasing and evolution of NVSS galaxies was presented by 
\cite{NusserTiwari2015}. Their best-fit bias function is comparable to Eq.~(\ref{bias}) up to $z \approx 1.5$ 
and grows slower at larger redshifts. We checked that this difference is insignificant for the purpose of this 
study.  For the additional bias due to non-Gaussianity see \cite{bias2000} and \cite{2008PhRvD..77l3514D}.

The expected $C_l$ for the NVSS catalogue are obtained using a modified version of 
the \class{} package \citep{DiDio:2013bqa}. The parameters for the best-fit cosmological model are taken from 
\cite{Planck2013CosmologicalParameters}. For the theoretical prediction of the angular 
two-point correlation we cut the Legendre series at $l_{\rm max} = 900$. We convinced ourselves 
that this cut-off is large enough to ensure numerical convergence of $w(\theta)$ at all angular scales 
considered in this work. 

\section{\label{sec:3} Angular two-point correlation of radio galaxies}

\subsection{Optimal estimation}
The angular two-point correlation could be estimated from \citep{Peebles1980}
\begin{equation}
 1+w(\theta)=\frac{DD}{N (N + 1)/2}\frac{\Omega}{\langle\delta\Omega\rangle},
\end{equation}
where $DD$ denotes the count of pairs at separation $\theta$ and $N$ denotes the total number of 
objects considered in the analysis. Thus $N(N+1)/2$ is the total number of possible pairs,
$\Omega$ is the solid angle of the survey, and $\langle\delta\Omega\rangle$ is the averaged solid angle of the 
ring $\theta$ to $\theta +\delta\theta$ within $\Omega$ for a randomly placed ring centre in $\Omega$. 

For our analysis we use the optimal estimator found by \cite{Landy:1993yu},
\begin{equation}
 w_{\rm LS}(\theta)=\frac{N_{\rm r}(N_{\rm r} + 1)}{N(N+1)} \frac{DD}{\overline{RR}} - 
 N (N_{\rm r} +1)\frac{\overline{DR}}{\overline{RR}}+1,
\end{equation}
where $\overline{RR}$ means the averaged pair count over a number of large random simulations, 
$\overline{DR}$ is the averaged data-random cross pair count for a number of large random simulations, and 
$N_{\rm r}$ denotes the number of sources in the random catalogues. 
It is necessary to clarify that the simulated random sources $R$ have to be identical for 
$\overline{RR}$ and $\overline{DR}$ to minimize the statistical uncertainty. 
Under the assumption that higher order correlations among galaxies can be ignored, 
the Landy-Szalay estimator has a ``Poisson error''
\begin{equation}
 \delta w_{\rm LS}(\theta) = \frac{1}{\sqrt{DD}}\frac{1+w_{\rm LS}(\theta)}{1+w_{\Omega}},
\end{equation}
where $w_{\Omega}$ is the mean of the two-point correlation function over the sampling geometry,
\begin{equation}
 w_\Omega=\int_\Omega G_p(\theta)w(\theta){\rm d}\Omega
\end{equation}
where $G_p(\theta)$ is a dimensionless geometric form factor which is equal to the fraction of unique cell pairs separated by distance $\theta\pm {\rm d}\theta/2$ \citep{Landy:1993yu}.

In \cite{Xia:2010yu} a pixel based estimator was used ($\Nside = 64$). We compared the 
Landy-Szalay estimator and the pixel based estimator at $\Nside = 64$ for the same flux threshold 
and mask. On angular separations above $5$ degrees, we found good agreement between 
both estimators. However, for smaller angular scales it seems that the pixel estimator is 
redistributing correlation from the smallest scales to intermediate scales. 
As for $\Nside = 64$ the distance between two neighbor pixel centers is as 
large as $0.9$ deg, it is not surprising that a pixellation error shows up at the few degrees scale. 

\subsection{Dipole correction}

The radio dipole signal is believed to result from our peculiar motion  \citep{1984MNRAS.206..377E} 
with respect to the cosmic rest frame of 
radio galaxies, because that the mean redshift of radio galaxies is above one. 
If this rest frame is the same as the CMB rest frame, then the dipole measured in the radio catalogue 
should agree with the CMB dipole measured by WMAP and Planck \citep{Jarosik:2010iu,Aghanim:2013suk}.
Previous studies\footnote{Note that the definition of the radio dipole amplitude in 
\cite{Blake:2002gx} differs from the rest of the literature by a factor of 2.}
\citep{Blake:2002gx,Singal:2011dy,Gibelyou:2012ri,Rubart:2013tx,Tiwarietal2014,Singal2014} measured 
the radio 
dipole for the NVSS catalogue. It is actually significantly larger than expected, by a factor of two to four depending 
on the details of the analysis. The origin 
of this dipole excess is currently unknown. One possibility might be that it is a combination of local 
large-scale structure \citep{Rubart:2014lia} and a kinetic component due to Doppler shift and aberration.

\begin{figure}
  \includegraphics[width=\linewidth]{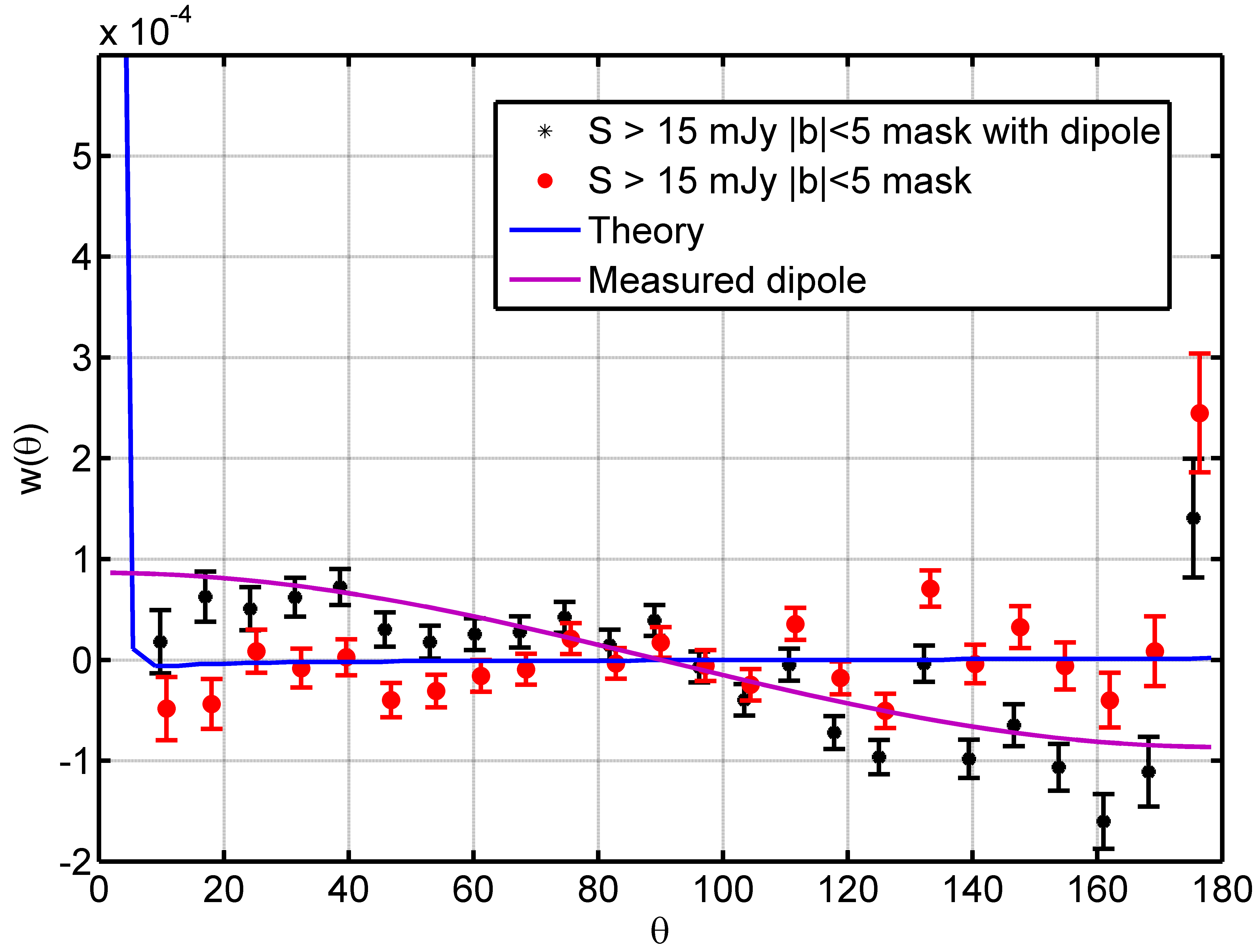}
  \caption{NVSS angular two-point correlations for a $5^\circ$ galactic latitude mask with and without dipole   
  correction.}
  \label{fig:NVSS-dipole}
\end{figure}

A local structure dipole, and the kinetic dipole as well, violate the assumption of statistical isotropy that is 
implicit in the way we estimate $w(\theta)$. We suggest that the dipole signal needs to be taken into 
account prior to the further correlation or power spectrum analysis. In our analysis 
the largest contribution to the angular two-point correlation of radio galaxies also comes from the dipole 
moment of the galaxy distribution, as can be seen in Fig.~\ref{fig:NVSS-dipole}.  A significant effect can 
also be observed for the extracted multipole moments at low $l$ (not shown here).

Utilizing \healpix, we find the radio dipole of the NVSS catalogue after masking with NVSS65 
(see table~\ref{tab:dipole}). The estimated dipole at our chosen flux density thresholds and sky 
coverage agrees with the estimates from the literature (see \cite{Rubart:thesis} for a recent summary).

\begin {table}
\begin{center}
\begin{tabular}{c | c | c | c | c }
 & $N$ & $d$ & $\alpha$ & $\delta$\\ \hline
$S>10$ mJy & 436,733 & $1.32\times10^{-2}$ & $142.70^{\rm o}$ & $30.47^{\rm o}$\\ 
$S>15$ mJy & 314,594 & $1.44\times10^{-2}$ & $153.44^{\rm o}$ & $-5.53^{\rm o}$\\ 
$S>25$ mJy & 200,092 & $1.83\times10^{-2}$ & $157.12^{\rm o}$ & $-15.10^{\rm o}$\\ \hline 
expected & & $0.46\times10^{-2}$ & $168^{\rm o}$ & $-7^{\rm o}$
\end{tabular}
\end{center}
\caption {NVSS dipole for various flux density thresholds, measured by means of \healpix\ at resolution 
$\Nside = 32$ after applying the NVSS65 mask. For comparison we quote the expected kinetic 
dipole for NVSS radio sources, based on the observed CMB dipole.}
\label{tab:dipole}
\end{table}

The standard dipole subtraction approach for pixelized maps is relatively straightforward.
First, we estimate the dipole amplitude and direction through a linear dipole estimator on the pixelized map.
Then we subtract the measured dipole contribution at each pixel. However, we do not use the pixel 
map to measure the correlation function, but rather extract it using the measured positions of all radio sources 
outside the mask. To achieve this, we have to include the effect of a dipole into the Landy-Szalay 
estimator. 

Thus we simulate random catalogues with the measured dipole, denoted by $R_{\rm d}$ below, and employ the following estimator, 
\begin{equation}
 w^{\rm d}_{\rm LS}(\theta)=\frac{N_{\rm r}(N_{\rm r} + 1)}{N(N+1)} \frac{DD}{\overline{RR}} - 
 N (N_{\rm r} +1)\frac{\overline{DR_{\rm d}}}{\overline{RR}}+ \frac{\overline{R_{\rm d}R_{\rm d}}}{\overline{RR}}.
 \label{eq:dLS}
\end{equation}
The dipole simulations $R_{\rm d}$ are achieved by the following procedure. 
First, we assign a uniform random number $t$ to each simulated object and then modify this number based on 
the angular separation $\psi$ between the object and dipole direction,
\begin{equation}
 t=\text{Random}[0,1)+ \frac{d}{2}\cos\psi
\end{equation}
Then we add the objects with $t \geq 0.5$ to the $R_{\rm d}$ catalogue and drop the others. Each random 
catalogue contains $N_{\rm r} = 10^6$ objects and we average over ten such catalogues.

The dipole modified estimator (\ref{eq:dLS}) makes use of the full position information of the sources and by 
simulating several large random catalogues, we minimize the uncertainty in $w^{\rm d}_{\rm LS}(\theta)$. 
The computational load is the disadvantage of this procedure.

We can now compare the results with and without dipole subtraction. We either corrected for 
the measured radio dipole (see table \ref{tab:dipole}) or for the CMB predicted radio dipole. 
The dipole contribution 
in the angular two-point correlation function can be seen from Fig.~\ref{fig:NVSS-dipole}. We find that 
the dipole has a significant effect and actually dominates the two-point correlation function at large angular 
scales above $\sim 10^\circ$. In the figure we account for the measured dipole. Considering just the 
CMB predicted dipole reduces the large-angle correlation, but leaves us with a 
residual dipole that could be due to a local structure and that is hard to predict without a much more 
detailed study. We therefore decided to correct for the 
measured NVSS radio dipole and also suppress the structure dipole in the theoretical prediction.

\subsection{Results}

We adopted the dipole subtracted Landy-Szalay estimator to measure the angular two-point correlations of the 
NVSS catalogue with thresholds of $15$ mJy and $25$ mJy for two different masks (NVSS65 and a constant 
latitude cut of the galactic plane). 

\begin{figure}
  \includegraphics[width=\linewidth]{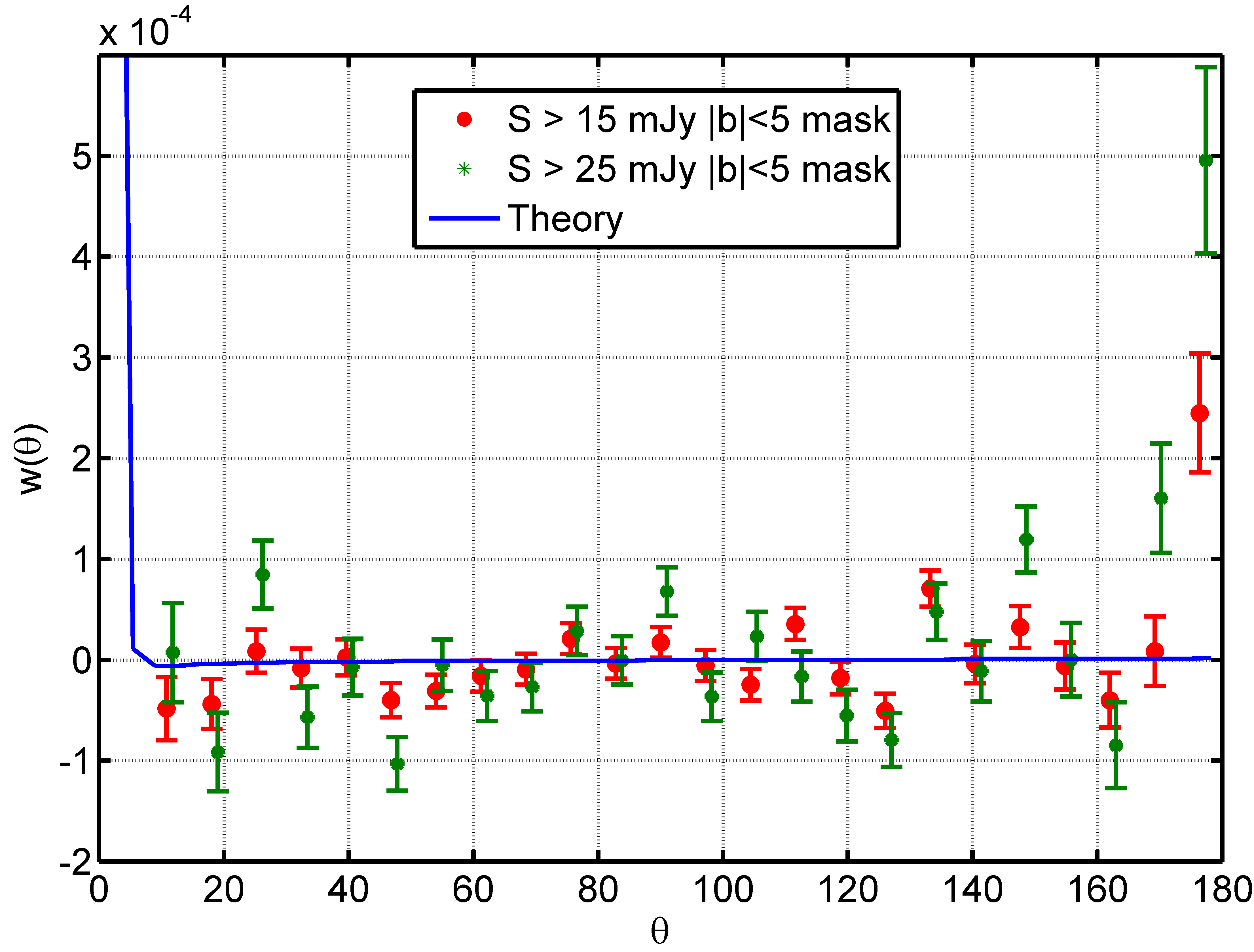}
  \includegraphics[width=\linewidth]{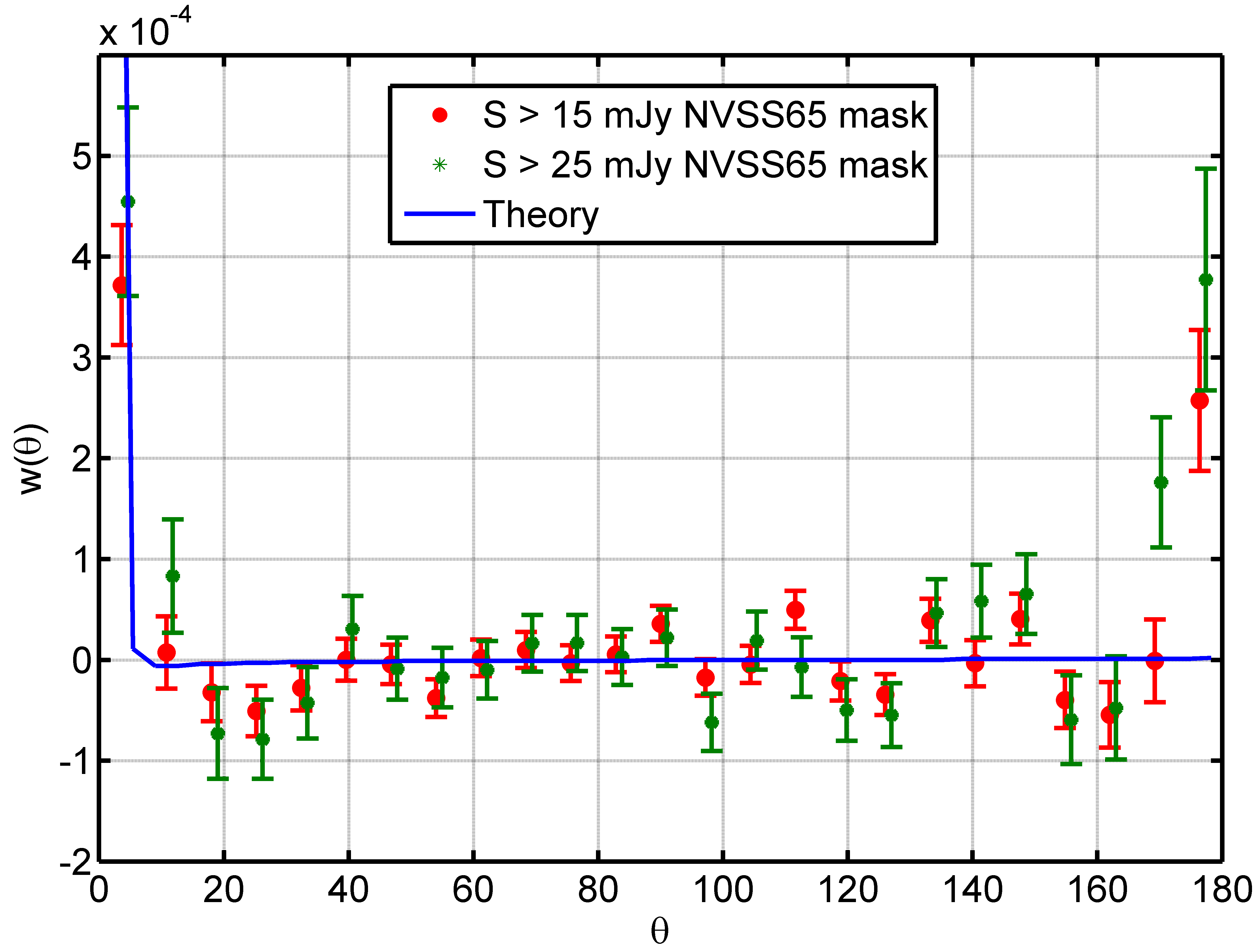}
  \caption{Angular two-point correlation function $w(\theta)$ from $5^\circ$ galactic latitude cut (top panel)
  and the NVSS65 mask (bottom panel). In both cases we include a dipole correction.}
  \label{fig:NVSS-W}
\end{figure}

The results agree with our expectation. After  eliminating the contribution from high rms 
noise and confusion pixels by means of the NVSS65 mask, the two-point correlation turns out to be less 
scattered, (see Fig.~\ref{fig:NVSS-W}). For the simpler constant latitude cut, the 
first data point at $\sim 4^\circ$ is above the plot range of the figure. The NVSS65 mask efficiently reduces 
the amount of correlation at scales of a few degrees and brings the 
measurement in agreement with the theoretical expectation of the best-fit cosmological model, and also 
observe a better agreement of the $15$ and $25$ mJy thresholded data sets after the NVSS65 mask has been 
applied. These findings are confirmed by the $\chi^2$-values shown in Table~\ref{tab:NVSS-chi2}. We infer 
that the new NVSS65 mask efficiently pushes the data points towards the theoretical prediction. 

\begin {table}
\begin{center}
\begin{tabular}{ c | c | c | c | c |}
 & \multicolumn{2}{c|}{ $|b|<5^{\rm o}$ }&\multicolumn{2}{c |}{ NVSS65 }\\ 
 \cline{2-5}
 & N & $\chi^2$ & N & $\chi^2$ \\ \hline 
$S>15$ mJy & 377,739 & $165.96$& 322,557 & $94.08$  \\ 
$S>25$ mJy & 240,872 & $222.49$& 205,103 & $99.80$  \\ 
\end{tabular}
\end{center}
\caption {$\chi^2$-test for $w(\theta)$ for 49 data points [excluding first bin ($0<\theta<3.6^\circ$)]. 
 We neglect the correlation between the data points.}
\label{tab:NVSS-chi2}
\end{table}

\begin{figure}
  \includegraphics[width=1\linewidth]{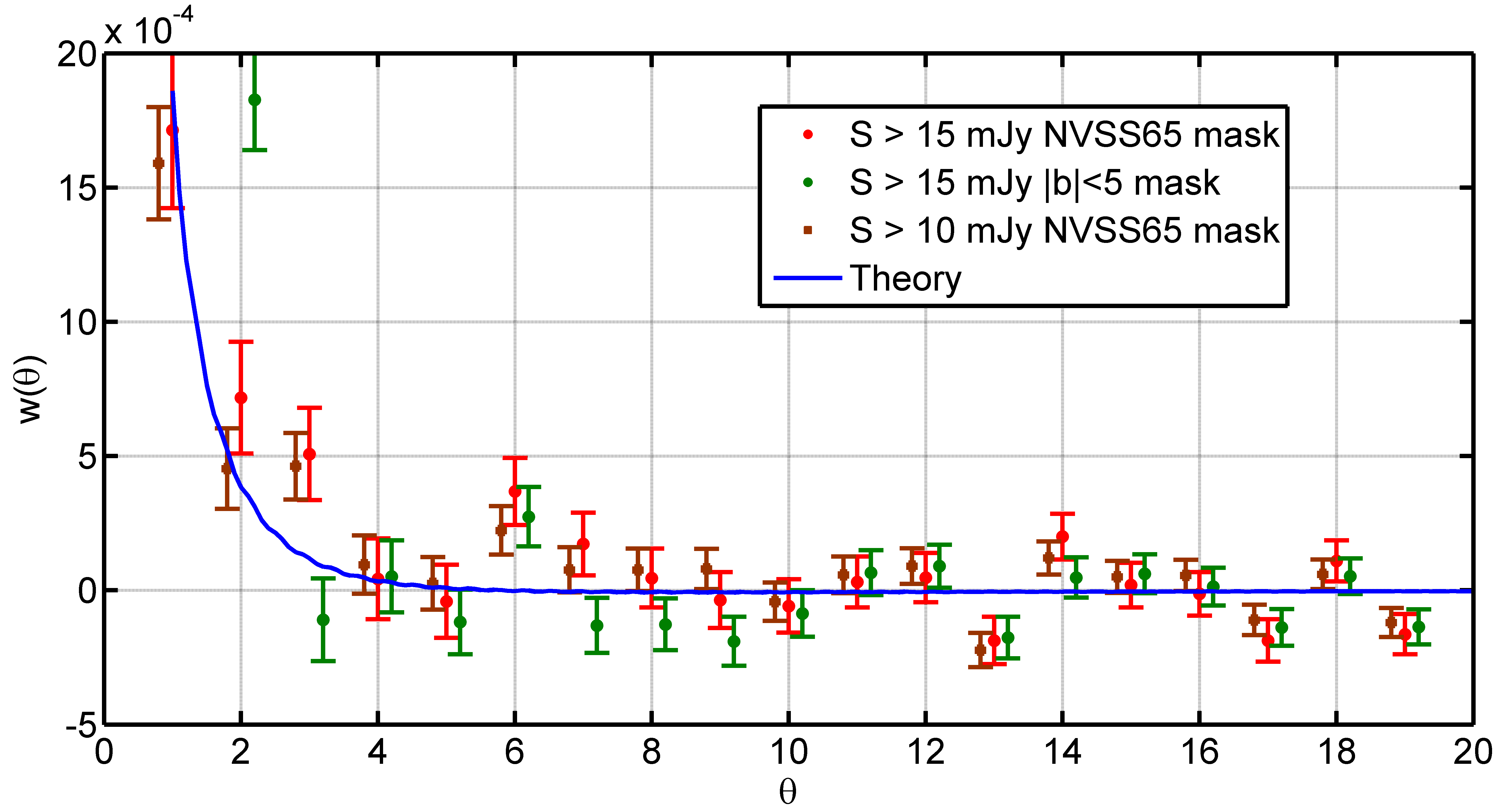}
  \caption{ Angular two-point correlation function at $1^\circ < \theta < 20^\circ$ from the 
  NVSS catalogue 
  with dipole correction for two different masks and two different flux thresholds.}
  \label{fig:small}
\end{figure}

Fig.~\ref{fig:NVSS-W} also shows that the measurement using the NVSS65 mask agrees with 
the theoretical prediction until $\theta \sim 90^\circ$. Above that angular scale the data appear to be more noisy.
The quite large correlations at the largest angular scales close to $180^\circ$ are surprising.

For an angular scale between $1^\circ$ and $20^\circ$, the 65\% sky coverage guarantees that the 
$w(\theta)$ estimator is averaged over a large number of independent sky patches.  
Artificial fluctuations caused by the survey or galactic foreground are suppressed in the average, and 
sidelobes and multicomponent source effects are expected to contaminate smaller angular scales 
(up to $0.6^\circ$). In this range, $w(\theta)$ is expected to be consistent with the theoretical prediction.

The result of an analysis at higher angular resolution for $\theta < 20^\circ$ is shown in Fig.~\ref{fig:small}. 
In order to suppress the shot noise contribution we now focus on the $S > 15$ mJy data set. 
We find that the NVSS65 mask improves the agreement with theoretical predictions considerably. 
At this angular scale, the most important effect of the NVSS65 mask is to remove spurious 
correlation at $\theta \leq 2$ deg.
We also observed that the NVSS65 mask increases the importance of the dipole subtraction at a separations 
of a few degrees. As \cite{Xia:2010yu} apply a flux threshold of $10$ mJy in their analysis,  we also show our
corresponding analysis for comparison, however as already discussed above, we think that this flux 
threshold is not a conservative choice.  A discussion of the cosmological consequences is given below. 

\begin{figure}
  \includegraphics[width=1\linewidth]{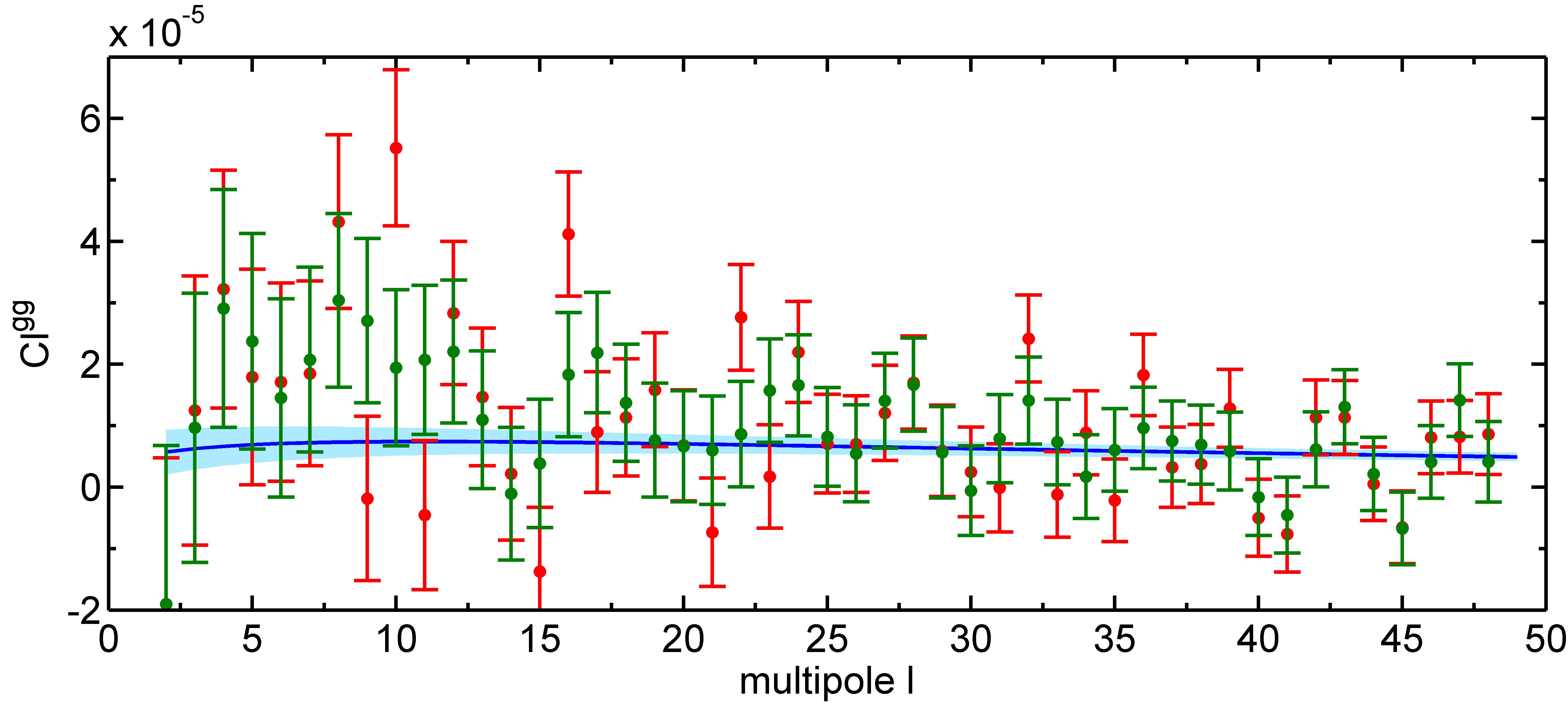}
  \caption{ NVSS angular power spectrum $C_l$ for $S>15$~mJy, dipole corrected and 
  NVSS65 masked.  $C_l$ is evaluated via a Legendre transformation from the angular two-point 
  correlation function (red). The extreme fluctuations are reduced, if the Legendre transformation 
  is limited to $\theta < 100$ deg (green).
  The solid line and the band around it show the theoretical prediction and its cosmic variance.}
  \label{fig:class}
\end{figure}

A complementary analysis to the angular two-point correlation function is to study the angular power spectrum.
One way of measuring $C_l$ is to do a Legendre transformation of the angular two-point correlation. The result 
of that transformation is shown in Fig.~\ref{fig:class}. Many of the $C_l$ turn out to have negative 
values, which shows that this measurement is quite noisy. Nevertheless, in the mean the $C_l$ seem to 
agree well with the theoretical expectations, with exception of a few multipoles with 
even $l$, most prominently the $l = 10$ mode. When restricting the Legendre transformation to the range 
$\theta < 100$ deg, assuming $w(\theta) = 0$ at $\theta > 100$ deg,  we find that the most extreme 
values, e.g. at $\ell = 10$ are suppressed.

\section{\label{sec:4}Discussion}

\subsection{Cosmological implications}


\begin{figure}
  \includegraphics[width=\linewidth]{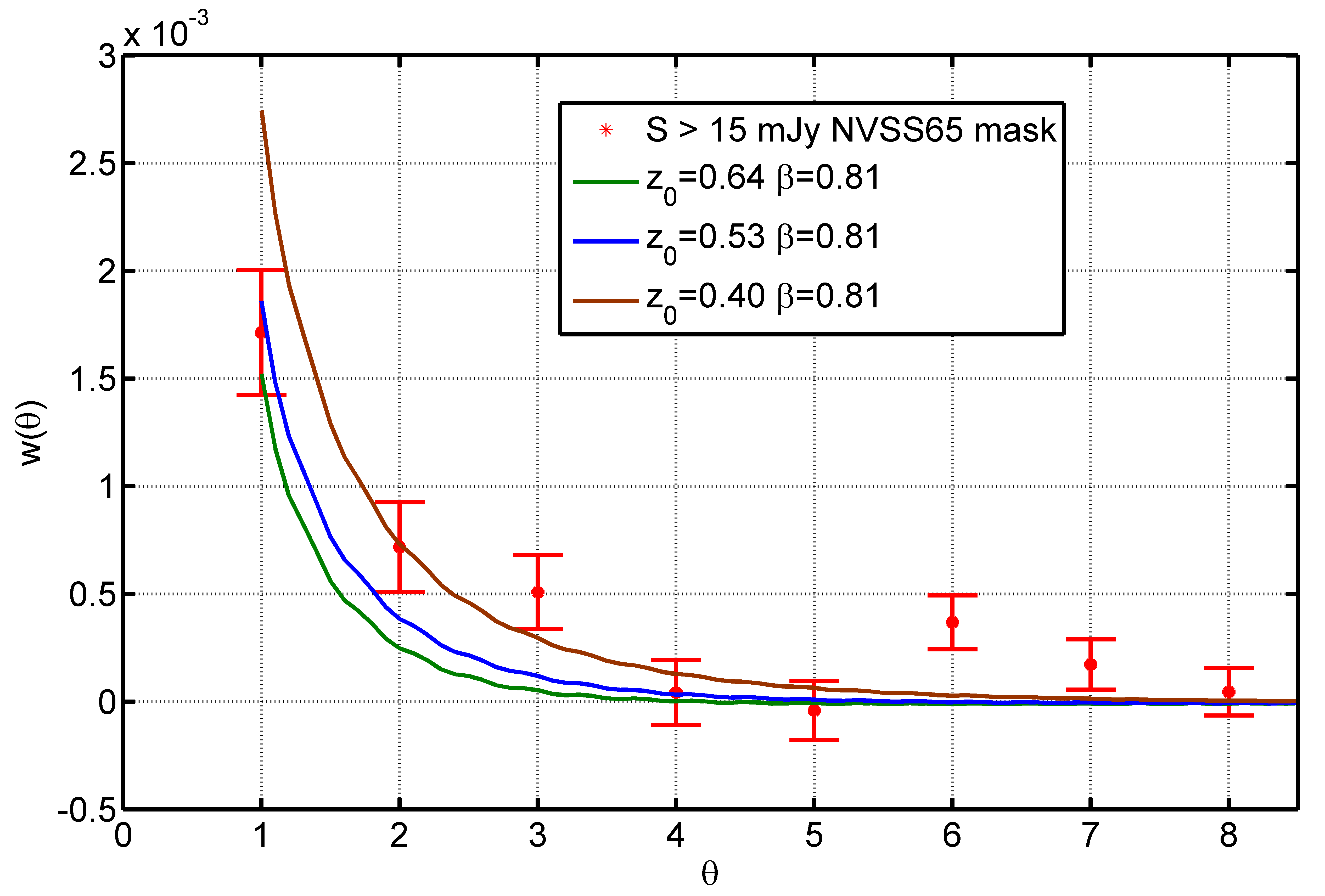}
  \caption{Angular two-point correlation function for different redshift distributions. }
  \label{fig:z0}
\end{figure}

We now turn to the cosmological implications and a discussion of our findings in comparison with previous 
studies. In the following we focus on the results obtained by means of the NVSS65 mask, including the dipole 
correction as discussed above and a lower flux threshold of $15$ mJy.  

A question of interest is the determination of the cosmological parameters of our minimal six-parameter 
cosmological standard model. For that purpose, the NVSS data alone cannot compete with high fidelity 
data from the CMB or from optical galaxy redshift surveys. Therefore it is not interesting to determine 
any of the six cosmological base parameters from a fit to NVSS.

  

It is important to note that the redshift distribution function also affects the NVSS angular 
two-point correlation at this angular scale. The Figure \ref{fig:z0} shows that 
the parameter $z_0$ changes its slope and and angular scale.
Our angular two-point correlation measurement seems to prefer low $z_0$ and $\beta$, which agrees with 
the results of \cite{Marcos-Caballero:2013yda}.

There is a clear degeneracy between cosmological parameters such as $H_0$, the redshift distribution 
parameters ($z_0$ and $\beta$) and the galaxy bias,
which limits our ability to use the most recent radio continuum surveys for cosmological parameter estimation. 
However, it is encouraging that the completely independent measurements of $H_0$ (Planck), 
$z_0$ and $\beta$ (CENSORS) provide a picture that seems to be consistent with 
$w(\theta)$ from NVSS. In order to extract the full science potential 
of upcoming radio surveys, a larger sample of redshifts must be measured.

To go beyond the set of the six base parameters, the 
small angle two-point correlation is also an interesting probe of primordial non-Gaussianity.
Previous results \citep{Xia:2010yu} claimed evidence of a primordial non-Gaussianity at the $3\sigma$ level.
The angular correlation function is supposed to vanish at around $4^\circ$ if the fluctuations are Gaussian, 
but their 
measurement of $w(\theta)$ showed a constant shift from zero at $1^\circ < \theta < 10^\circ$, which they 
attributed to the effect of a primordial non-Gaussianity. However, adapting our procedure of masking, 
dipole correction and optimal estimation of $w(\theta)$, we do not see this shift. Our result is 
consistent with primordial Gaussianity, as is shown in Fig.~\ref{fig:small}. 
Turning that into a precise limits on $f_{\rm nl}$ is beyond the scope of this work.

A primordial non-Gaussianity would also lead to an increase of angular power $C_l$ at low multipoles 
\citep{Xia:2010yu,2010CQGra..27l4011D}. The angular power spectrum obtained via a Legendre 
transformation is shown in Fig.~\ref{fig:class}. One could be tempted to conclude that we do find an 
increase of power for low multipoles, similar to 
\cite{Blake2004a} and \cite{Marcos-Caballero:2013yda}. However, it should be noted that our method 
to estimate the angular power spectrum is not optimal and appears to be noisy. On top, most of the 
excess power is in parity even modes, primordial (local) non-Gaussianity would not care about parity.

We tried to understand what causes for the removal of the apparent non-Gaussianity in 
the correlation function, but it turns out that it is impossible to attribute it to a single effect. 
As already discussed in detail above, the essential steps are, 
using the Landy-Szalay estimator (to avoid pixellation errors), using a larger mask, and removing 
the dipole contribution.

\subsection{Residual systematics} 

The theoretical prediction based on the $\Lambda$CDM model suggested $C_l \approx 5\times 10^{-6}$. Our 
measurements closely agree  with that prediction, with some exceptions. The quadrupole cannot be detected
at any significant level and the power at $l = 10$ is an order of magnitude larger than expected. 
However, at larger $l$, up to $l \sim 60$, the $C_l$ are consistent with the 
theoretical prediction, which is in agreement with previous analyses of the ISW effect \cite{refId0}. 
Higher multipole moments are noisy and statistically consistent with zero.

We finally discuss the angular scales at $\theta > 20^\circ$ and look in more detail at the corresponding 
multipole moments up to $l = 10$. For a simple galactic isolatitude cut, we find significantly more power at 
low multipole moments, with $l = 4$ being the dominant mode (not discussed here). The essential way to get rid 
of this extra power is to take direction dependent systematic effects into account in the NVSS catalogue. 
Our NVSS65 mask allows us to reduce the $l=4$ mode and to recover an overall flat angular power spectrum.    
However, even with the NVSS65 mask, some of the $C_l$s are one order of magnitude larger than the 
prediction at $l<10$. Perhaps because of remaining surface density fluctuations at 
different declinations.  

\begin{figure}
  \includegraphics[width=1\linewidth]{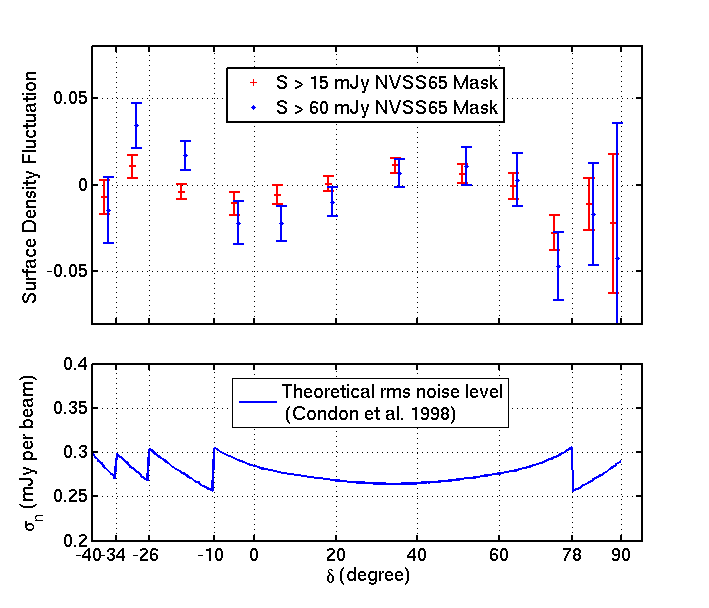}
  \caption{Top panel: Surface density fluctuation ($\Delta\sigma/\overline{\sigma}$) of NVSS
   sources at $S>60$ mJy and $S>15$ mJy after applying the NVSS65 mask.  
   Bottom panel: The theoretical rms noise level of the NVSS catalogue 
   from \cite{Condon1998}. The VLA geodetic latitude is $34^\circ 04'44''$.}  
   \label{fig:strip}
\end{figure}

We find a clear anti-correlation between the surface density and the theoretical rms noise of the NVSS catalogue, 
Fig.~\ref{fig:strip}. The declination dependence of the theoretical rms noise fluctuations at 
$-40^{\circ}<\delta<-10^{\circ}$ is due to changes in the snapshot integration time.
When the VLA points to the horizon, the effect of the projection of the uv plane combined with ground 
and ionospheric noise decreases the effective signal-to-noise ratio of the survey. As can be seen in 
Fig.~\ref{fig:strip} this effect is not limited to the faintest sources, but is also there for brighter sources at 
$S > 60$ mJy. Thus, only the most extreme influence of the direction dependent systematics can be cured by 
means of a lower flux threshold. 

\begin{figure}
  \includegraphics[width=1\linewidth]{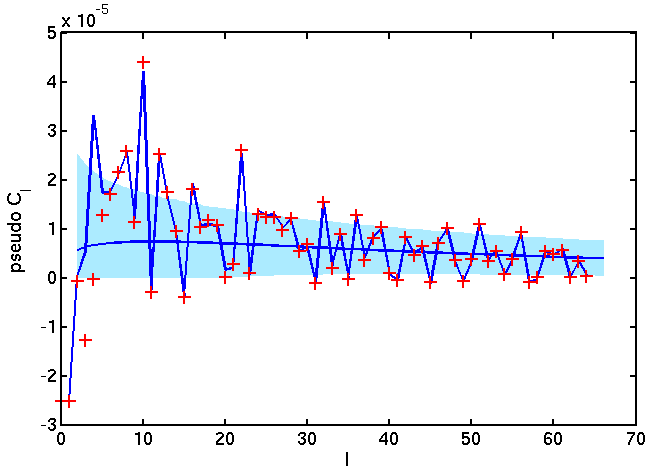}
  \caption{ The \healpix{} pseudo-$C_l$ of the NVSS catalogue with NVSS65 mask for 
  $S>15$ mJy and radio dipole subtracted. 
  The zig-zac line shows the standard normalization. The crosses show the declination normalization result.
  The smooth solid line shows the theoretical prediction and the band around it includes shot noise and 
  cosmic variance.}
  \label{fig:Healpix}
\end{figure}

To further probe whether these direction dependent effects affect the low-$l$ multipoles, we investigated the 
angular pseudo-power spectra, which are obtained through a decomposition into  spherical harmonics on 
the pixelized galaxy number count. For this we used \healpix. In figure~\ref{fig:Healpix} we show estimates 
of the pseudo-$C_l$ for two different normalizations. In the first case, we normalize the pixel count with the 
mean pixel count of the full survey $\bar{n}$, i.e. 
\begin{equation}
 x_i= \frac{n_i-\bar{n}}{\bar{n}} \;.
\end{equation} 
The second case is motivated by \cite{2006MNRAS.365..891V}. We divide the map along the declination, based 
on durations of the snapshots, and subtract the mean pixel count $\bar{n}_{\delta_i}$ at declination $\delta_i$,
\begin{equation}
 x_i= \frac{n_i-\bar{n}_{\delta_i}}{\bar{n}} \;.
\end{equation}
This procedure was also used in the Planck analysis of the ISW effect \citep{planckISW,planckISW15}. 

This second procedure, erases the source density fluctuations along the declination direction. As can be seen in 
Fig.~\ref{fig:Healpix}, the $l=3,4,5$ multipoles are significantly reduced by the declination 
mean subtraction, which strongly implies that the $l=4$ mode fluctuations are caused by the direction dependent 
noise of the NVSS catalogue. On the other hand, the large $l=10$ mode is not affected by that normalization at 
all and it is not clear why it exceeds the expectation by an order of magnitude. Using this declination 
normalization for a cosmological analysis, such as \cite{2006MNRAS.365..891V}, \cite{Fernandez-Cobosetal2014} 
or \cite{planckISW},
certainly suppresses fluctuations on large scales, but the procedure cannot distinguish direction dependent 
effects from real fluctuations and it is very hard to assign an error estimate to it.

\begin{figure}
  \includegraphics[width=1\linewidth]{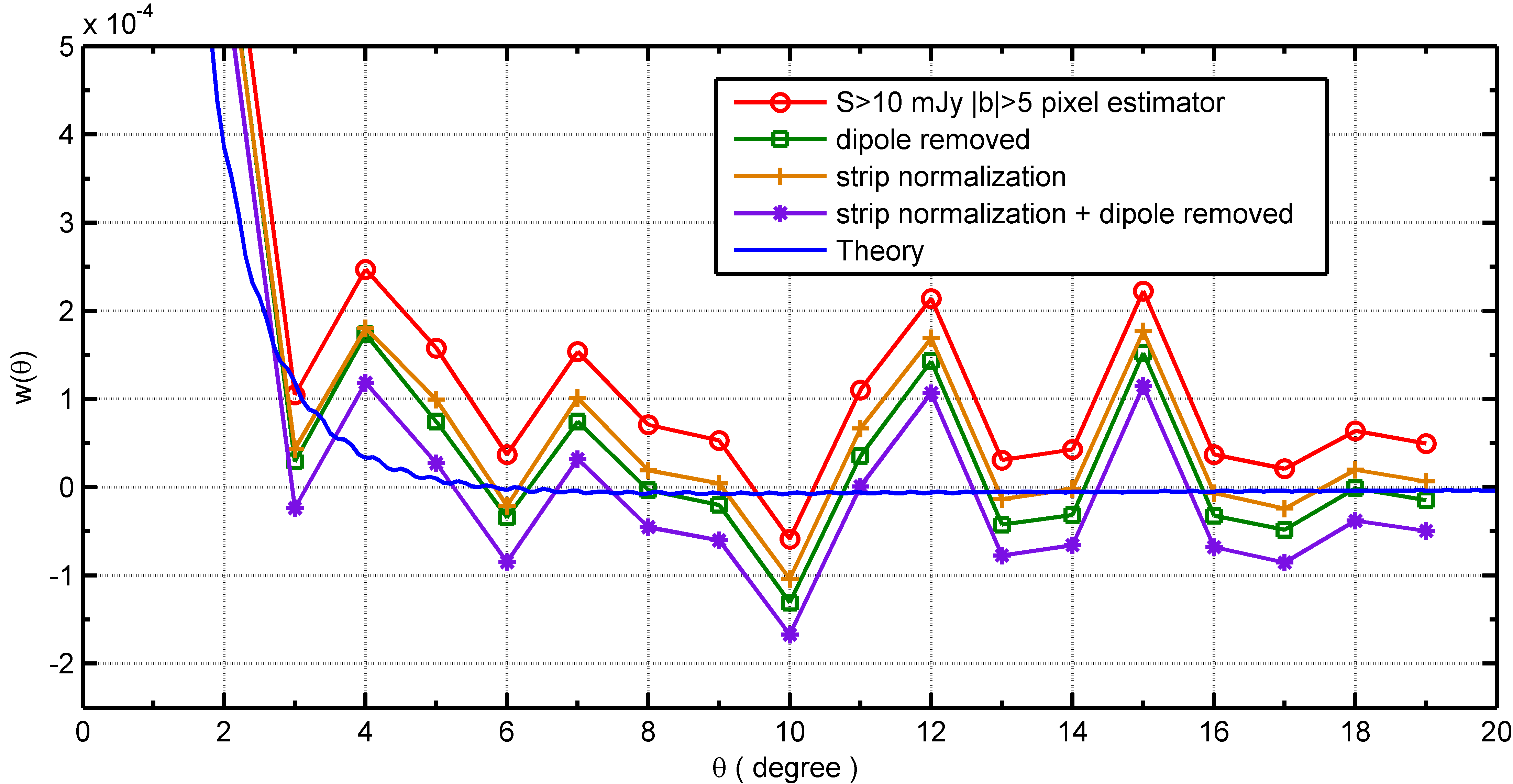}
  \caption{Pixel based estimate of the angular correlation function of the NVSS catalogue for 
  $S > 10$ mJy with $|b|<5$ mask (red line). The effects of dipole removal (green) and declination 
  normalisation (brown) are shown for comparison. The purple line shows the result using 
  a declination normalization and subsequent dipole subtraction.}
  \label{fig:normalization_vs_dipole}
\end{figure}

To analyse this effect in more detail, we took the $S > 10$ mJy, $|b|<5$ degree map before dipole subtraction 
and used the same pixel estimator as in \cite{Xia:2010yu} (including only full pixels) as a baseline in 
Fig.~\ref{fig:normalization_vs_dipole}. This baseline seems to have some excess power.  
We found that both the strip normalization and the dipole subtraction give a substantial reduction and 
almost identical results. We thus think that it would be better to just subtract the dipole and to use a more 
conservative mask or carefully modeling this systematic effect, 
instead of setting all fluctuations between declination strips to zero.

In addition, we note a curious observation.
Looking carefully at Fig.~\ref{fig:class}, we observe a parity asymmetry in the $C_l$: the power of 
even multipoles $l = 4, 8, 10, 12, 16, 22, 24, 32...$ is higher than that of their odd neighbours. This even 
parity preference can also be seen because $w(180^{\circ}) > 0$. A similar effect was observed in the 
CMB angular power spectrum, where the parity asymmetry is the opposite \cite{Kim:2010gf}. There the 
odd $l(l+1)C_l/2\pi$s are larger than the even ones.  

Our results further suggest that $w(\theta)$ from full-sky radio continuum surveys can be used to constrain 
cosmological parameters at an epoch that is barely accessible to other probes. Planned and upcoming 
surveys with instruments like LOFAR, ASKAP, MeerKAT and finally SKA will allow us to reduce the 
shot noise and to increase angular resolution and sensitivity, while covering all sky, extending the 
studies to several frequency bands (our discussion here is limited to $1.4$ GHz), and improving the 
control of systematic effects. For more details see \cite{Raccanelli:2011pu} and \cite{Jarvis:2015asa} 
and references therein.

\section{\label{sec:6}Conclusion}

In this paper we revisited the angular two-point correlation function $w(\theta)$ and angular power spectrum 
$C_l$ from the NVSS catalogue of radio galaxies. The focus of our work was to investigate systematic 
effects in the NVSS catalogue. In order to minimize the contribution of these effects in the cosmological analysis, 
we provide a new NVSS mask with 64.7\% percent of sky, called NVSS65. We also find that it is essential to 
account for the radio dipole and to use an optimal estimator. We found that our mask significantly 
improves the $\chi^2$ value of the angular two-point correlation function on all angular scales. 
For angular scales between $1$ and $20$ degrees, $w(\theta)$ agrees with the flat $\Lambda$CDM 
model without introducing primordial non-Gaussianity. 

Thus we have shown that to fully explore the cosmological potential of continuum radio surveys, one has to
understand and investigate the systematic effects related to flux calibration, especially direction dependent 
effects of the calibration. In addition, the effect of the cosmic radio dipole affects the reconstruction of 
higher multipole moments and the attempts to measure or constrain primordial non-Gaussianity.   

To obtain an improved upper limit on $f_{\rm nl}$ or to constrain other cosmological parameters 
at a redshift of about unity is beyond the scope of this work, as it would need an extensive study of the 
uncertainties coming from our understanding of the density, luminosity and bias evolution of radio 
galaxies. However, our analysis shows that the radio sky is in remarkably good agreement with the 
standard model of cosmology according to Planck. It will also be interesting to improve the 
ISW analysis of the Planck-NVSS cross-correlation by means of the new NVSS65 mask, and to 
include a radio dipole correction, as well as a higher flux threshold. 

\begin{acknowledgements}
We thank  Anne-Sophie Balleier, Daniel Boriero, Bin Hu, Dragan Huterer, Hans-Rainer Kl\"ockner, 
Alvise Raccanelli, and Matthias Rubart for valuable comments and discussions. 
We acknowledge financial support from Deutsche Forschungsgemeinschaft (DFG) via the Research 
Training Group ``Models of Gravity'' (RTG 1620). We are grateful for the possibility of performing the 
numerical computations on the Nordrhein-Westfalen state computing cluster at RWTH Aachen.
We acknowledge the use of the NVSS catalogue \cite{Condon1998}, provided by the NRAO.
This work made extensive use of the \healpix{}~\cite{Gorski:2004by} and the \class{}~\cite{Lesgourgues:2011rh,DiDio:2013bqa} packages.
\end{acknowledgements}


\bibliographystyle{aa}
\bibliography{NVSS.bib}

\end{document}